\documentclass[twocolumn,showpacs,prX,nofootinbib,nobibnotes]{revtex4}

\usepackage{graphicx}
\usepackage{dcolumn}
\usepackage{bm}

\usepackage{anysize}
\usepackage{graphicx}
\marginsize{1.8 cm}{1.5 cm}{0.5 cm}{3.0 cm}

\usepackage{graphicx}
\usepackage{dcolumn}
\usepackage{bm}
\usepackage{amsmath}
\usepackage{amssymb}
\usepackage{mathrsfs}
\usepackage[latin1]{inputenc}
\usepackage{amsmath}

\newcommand{\dd}{{d}}
\newcommand{\p}{\partial}
\newcommand{\e}{\varepsilon}
\newcommand{\sd}{Schr\"{o}dinger }

\newcommand{\U}{\mathcal{U}}

\newcommand{\R}{\mathbb{R}}
\newcommand{\tr}{{\rm Tr}}

\begin{document}


\title{Quantum Multiobservable Control}

\author{Raj Chakrabarti}
\affiliation{Department of Chemistry, Princeton University,
Princeton, New Jersey 08544, USA} \email{rajchak@princeton.edu}

\author{Rebing Wu}
\affiliation{Department of Chemistry, Princeton University,
Princeton, New Jersey 08544, USA}

\author{Herschel Rabitz}
\affiliation{Department of Chemistry, Princeton University,
Princeton, New Jersey 08544, USA}

\date{27 April 2008}

\begin{abstract}

We present deterministic algorithms for the simultaneous control
of an arbitrary number of quantum observables. Unlike optimal control approaches
based on cost function optimization, quantum multiobservable tracking control (MOTC)
is capable of tracking predetermined homotopic trajectories to target expectation values
in the space of multiobservables. The convergence of these algorithms is facilitated by
the favorable critical topology of quantum control landscapes. Fundamental properties
of quantum multiobservable control landscapes that underlie the efficiency of MOTC,
including the multiobservable controllability Gramian, are introduced. The effects of multiple
control objectives on the structure and complexity of optimal fields are examined.
With minor modifications, the techniques described herein can be applied to general
quantum multiobjective control problems.

\end{abstract}


\maketitle

\tableofcontents

\section{Introduction}\label{intro}

The optimal control of quantum dynamics (QC) is receiving increasing
attention due to widespread success in laboratory experiments and numerical
simulations across a broad scope of systems. With these promising
results it becomes imperative to understand the reasons for success
and to develop more efficient algorithms that can increase objective
yields, especially for more complex objectives.

Quantum optimal control problems studied to date fall into two major
classes: 1) control of the expectation values of single quantum observables or
pure states \cite{Assion1998,HoRab2007a,RabMik2004,PeiDah1988}; 2)
control of quantum dynamical transformations \cite{Kosloff2002,Kosloff2003}.
The former has been implemented in both simulations and experiments, whereas the
latter has been approached predominantly through numerical studies, due to the
expense of quantum process tomography. An important third class of
QC problems, which lies between the latter two problem types,
is the control of arbitrary numbers of quantum observables.
To date, few effective techniques - either experimental or numerical - have
been reported for multiobservable quantum control \cite{Weber2007,Wolf2007}.

The typical experimental approach to controlling single quantum observables
is to maximize (or minimize) a control objective functional, such as the
expectation value of an observable operator, using stochastic search
algorithms. A common technique is to randomly sample single
observable expectation values at various points over the control
landscape and use genetic algorithms (GA) to update the control
field \cite{Levis2001}.  However, the utility of such stochastic
search algorithms needs careful consideration for more complex QC problems like
multiobservable control. A recent work \cite{Wolf2007} examined the application of multiobjective
evolutionary algorithms (MOEA), which do not rely on a control objective functional, to a
two-observable quantum optimal control problem. While the MOEAs displayed promising improvements
over GAs, they scale more poorly than objective function-based algorithms \cite{Deb2002}.
Moreover, both GAs and MOEAs are limited in their efficiency by the fact that they do not make
use of prior knowledge pertaining to the structure of the search landscape. As such, there
remains a need for the development of both numerical and experimental control strategies
tailored for the problem of multiobservable quantum control.

Recently, significant strides have been made towards establishing a
foundation for the systematic development of efficient optimal control
algorithms for more complex QC problems based on the observation that the
landscape traversed by search algorithms in the optimization of quantum controls is not
arbitrarily complicated, but rather possesses an analytical
structure originating in the geometry of quantum mechanics
\cite{Raj2007}. Prior work established important features of these landscapes, in
particular their critical topologies \cite{RabMik2004,WuPech2008}. The critical points of a
control landscape correspond to locally optimal solutions to the
control problem. The landscapes corresponding to problems 1) and 2)
were shown to have a set of critical submanifolds of measure zero in the search domain,
and to be essentially devoid of local traps, i.e., the vast majority of
local suboptima are saddles, facilitating the convergence of local
search algorithms.

The underlying monotonicity of quantum control landscapes has increased
interest in deterministic quantum control optimization algorithms. A recent
experimental study \cite{Roslund2007} demonstrated at least a two-fold improvement
in optimization efficiency through the use of local gradient algorithms rather than
a GA. Such experimental gradient-based
 search algorithms, enabled by the favorable
critical topology of QC landscapes, may in fact be essential for high-precision quantum
multiobservable control. Still, any optimal control theory (OCT) strategy based on
optimizing a control cost functional may not be ideal for  multiobservable control problems,
where the most effective combination of observable expectation values is not always clear at the outset.
A more powerful deterministic algorithm would  offer the ability to explore arbitrary
trajectories across the control landscape to identify desirable solutions.

In order to address this need, we develop in this paper the general theory
of quantum multiobservable tracking control (MOTC), a control strategy that seeks
to drive a quantum system along homotopic paths to desired multiobservable expectation values.
This algorithm is motivated by the so-called continuation methodology for multiobjective
optimization \cite{Hillermeier2001}, which is a vector optimization technique based on the
principles of differential geometry that serves as an alternative to multiobjective stochastic
optimization. The paper is organized as follows. In Section II, we derive analytical results
characterizing the gradient flows of quantum multiobservable control cost functionals, examining
the factors that govern the convergence of these flows, and showing that they follow paths
(in the unitary group) that are highly Hamiltonian-dependent. In Section III, we review the theory of unitary propagator tracking control, highlighting its algorithmic advantages compared to scalar
objective optimization, as well as its stringent requirements for the regularity of control fields. In
Sections IV and V, we develop the related theory of multiobservable tracking
control and introduce the MOTC Gramian matrix, which characterizes the ability of the search
algorithm to move in arbitrary directions in multiobservable space from a given point on the
control landscape. In Section VI, we describe the methods employed in the
numerical implementation of multiobservable tracking and then, in Section
VII, we present numerical illustrations of MOTC in the case of several model systems,
examining the effect of multiple observable constraints on the structure and complexity
of the optimal controls. Finally, in Section VIII, we draw general conclusions and discuss
future directions.

\section{Quantum multiobservable control gradient flows}\label{gradientflows}

A generic quantum optimal control problem can be written
\cite{Raj2007}:
\begin{equation}\label{OCT}
\max_{\e(t)}~~\Phi(U(T))
\end{equation}
where $U(T)$ is an implicit functional of $\e(t)$ via the \sd
equation for the unitary propagator
$$\frac{\dd U(t)}{\dd t}=-
\frac{\imath}{\hbar}H(\e(t))U(t), ~~~U(0)=I_N.$$ Here $H$ denotes the
total Hamiltonian and $\e (t)$ is the time-dependent control field.
Solutions to the optimal control problem correspond to $\frac{\delta
\Phi}{\delta \e(t)} = 0$ for all $t\in[0,T]$. The objective function $\Phi$
 can take various forms. The most common form of $\Phi$ is the expectation value of an
observable of the system:
\begin{equation}\label{obs}
\Phi(U(T)) = \tr(U(T){\rho(0)}U^{\dag}(T)\Theta),
\end{equation}
where $\rho(0)$ is the initial density matrix of the system and
$\Theta$ is the Hermitian observable operator whose expectation
value is to be maximized \cite{Mike2006a}.

A natural objective function for multiobservable control is a
positively weighted convex sum of individual observable objectives,
i.e.,
\begin{equation}\label{multi}
\Phi_{M}(U) = \sum_{k=1}^m \alpha_k \Phi_{k}(U), \quad \alpha_k > 0,
\end{equation}
where $\Phi_{k}(U)= \tr(U\rho(0)U^{\dag}\Theta_k), \quad
k=1,2,\cdots,m$. The goal of a multiobjective optimization
problem may be to maximize the expectation values of all
observables, i.e.,
\begin{equation*}
~\vec{\Phi}(U(T)) = \{\Phi_1(U(T)),\cdots,\Phi_M(U(T))\}.
\end{equation*}
Alternatively, the goal may be to target observable expectation
values $\chi_k$, $k=1,\cdots,m$, in which case objective function (\ref{multi}) can be replaced by
\begin{equation}\label{multi2}
\Phi_{M}'(U) = \sum_{k=1}^m \alpha_k|\Phi_{k}(U)-\chi_k|^2, \quad
\alpha_k > 0.
\end{equation}

In this section, we examine the factors that affect the efficiency
of algorithms that optimize scalar objective functions of the form
(\ref{obs}) or (\ref{multi}), with a specific focus on gradient
algorithms. We begin by writing expressions for the gradients of
these functionals. An infinitesimal functional change in the Hamiltonian
$\delta H(t)$ produces an infinitesimal change in the dynamical
propagator $U(T)$ as follows:
\begin{equation}\label{du}
\delta U(T) = - \frac {\imath}{\hbar} \int_0^T U(T)U^\dag(t) \delta H(t)
U(t)\dd t.
\end{equation}
The corresponding change in $\Phi$ is then given by
$$\delta \Phi = -\frac {\imath}{\hbar} \int_0^T
\tr\Big([\Theta(T),U^{\dag}(t)\delta H(t)U(t)]\rho(0)\Big)\dd t,$$
where $\Theta(T) \equiv U^{\dag}(T){\Theta}U(T)$. In the special
case of the electric dipole formulation, the Hamiltonian becomes
\begin{equation}\label{ham}
H(t) = H_0 - \mu \cdot \e(t)
\end{equation}
where $H_0$ is the internal Hamiltonian of the system and $\mu$ is
its electric dipole operator. In this situation, the gradient of $\Phi$ is \cite{HoRab2007a}:
\begin{equation}\label{grad1}
\frac{\delta \Phi}{\delta \e(t)} =
-\frac{\imath}{\hbar}\tr\Big(\left[\Theta(T),\mu(t)\right]\rho(0)\Big),
\end{equation}
where $\mu(t)= U^{\dag}(t)\mu U(t)$, and the gradient for $\Phi_M$
is
\begin{equation}\label{gradmult}
\frac{\delta \Phi_M}{\delta \e(t)} =
-\frac{\imath}{\hbar}\sum_{k=1}^m \alpha_k
\tr\Big(\left[\Theta_k(T),\mu(t)\right]\rho(0)\Big).
\end{equation}

The flow trajectories followed by these gradients are the solutions to the
differential equation
\begin{equation}\label{Egrad}
\frac{\p\e_s(t)}{\p s}= \gamma \frac {\delta \Phi_{(M)}}{\delta
\e_s(t)}
\end{equation}
where $s > 0$ is a continuous variable parametrizing the algorithmic
time evolution of the search trajectory, and $\gamma$ is an
arbitrary positive constant that we will set to 1. Prior work
\cite{Raj2007} demonstrated that under the assumption of linear
independence of the elements of the time-evolved dipole operator
$\mu(t)$, the landscape for objective function (\ref{obs}) contains
no local maxima, thus ensuring that the gradient flow (\ref{Egrad})
cannot be trapped. Furthermore, it was shown that the critical set of this objective
function consists of submanifolds of Lebesgue measure zero in
$\U(N)$ --- indicating that the likelihood of encountering a
suboptimal critical point is essentially null. Since equation (\ref{gradmult}) is the
gradient of the expectation value of a single observable $\Theta_M
 = \sum_{k=1}^m \alpha_k \Theta_k $, it follows
that its flow will also converge to the global optimum of
the objective function and share the above favorable landscape
features. These features indicate that gradient-based algorithms may also
be effective for multiobservable control optimization.

The development of such deterministic search algorithms for quantum multiobservable
control is especially important because of the aforementioned difficulties in sampling
multiobjective landscapes with stochastic techniques.
However, there are two characteristics of the simple gradient flows (\ref{Egrad})
that could be improved to render them more efficient in searching multiobservable control landscapes.
First, the convergence rate of gradient flow control optimization is highly Hamiltonian-dependent.
To explicitly isolate the Hamiltonian-dependent contribution to the search dynamics, consider first
the gradient flow of $\Phi_M$ on the unitary group $\U(N)$, which is given by \cite{WuMike2008}:
\begin{eqnarray}\label{Uflowmult}
\frac{\dd V_s}{\dd s} &=& \nabla \Phi_M(V_s)=\sum_{k=1}^m\alpha_k[\Theta_k,V_s\rho(0) V_s^{\dag}]V_s\\
&=& [\Theta_M,V_s\rho(0) V_s^{\dag}]V_s,
\end{eqnarray}
with $V \in \U(N)$. $\nabla \Phi_M(\cdot)$ denotes the gradient of the objective function on $\U(N)$, where the Riemannian metric is defined by the inner product
$$\langle X, Y \rangle \equiv \tr(X^{\dag}Y),$$ for any $X$ and $Y$ in the tangent space $T_V\U\equiv \{VB \mid B^{\dag} = -B\}$ of $\U(N)$ at $V$.

We are interested in the relationship between the
paths followed by the gradient flow (\ref{Egrad}) on $\e(t)$ and that (\ref{Uflowmult}) on $\U(N)$. The gradient function on $\e_s(t)$ is related to the gradient on $\U(N)$ through
\begin{eqnarray}\label{chain}
\frac{\delta \Phi_M}{\delta \e_s(t)}&=& \tr\left\{\nabla\Phi_M(U_s(T))\frac{\delta U^{\dag}_s(T)}{\delta \e_s(t)}\right\}\\
&=&\sum_{p,q=1}^N\frac{\p \Phi_M}{\p (U_s(T))_{pq}}\frac{\delta (U_s(T))^*_{pq}}{\delta
\e_s(t)}.
\end{eqnarray}
Now suppose that we have the gradient flow of $\e_s(t)$ that
follows (\ref{Egrad}) and let $U_s(T)$ be the projected trajectory
on the unitary group $\U(N)$ of system propagators at time $T$,
driven by $\e_s(t)$. The algorithmic time derivative of $U_s(T)$
is then
\begin{equation}\label{Us}
  \frac{\dd (U_{s}(T))_{ij}}{\dd s}=  \int_0^T \frac{\delta (U_{s}(T))_{ij}}
  {\delta \e_s(t)}
  \frac{\partial \e_s(t)}{\partial s} \dd t
\end{equation}
which, combined with (\ref{Egrad}) and (11) , gives
\begin{equation}\label{dot Us}
     \frac{\dd (U_{s}(T))_{ij}}{\dd s}=\int_0^T \frac{\delta (U_s(T))_{ij}}
     {\delta \e_s(t)}\sum_{p,q=1}^N\frac{\p \Phi_M}{\p (U_{s}(T))_{pq}} \frac{\delta (U_{s}(T))_{pq}^*}{\delta \e_s(t)}
     \dd t.
\end{equation}
To obtain expressions for the $\frac{\delta (U_s(T))_{ij}}{\delta \e_s(t)}$, note that
in the electric dipole formulation, equation (\ref{du}) becomes
\begin{equation}\
\delta U(T) = - \frac {\imath}{\hbar} \int_0^t U(T)U^\dag(t) \mu
U(t)\delta \e(t)\dd t.
\end{equation}
Taking the derivative of this expression with respect to a functional change in the
control field, we get
$$\frac{\delta U(T)}{\delta \e(t)}=-\frac{\imath}{\hbar}U(T)U^{\dag}(t)\mu U(t):=-\frac{\imath}{\hbar}U(T)\mu(t).$$
It is convenient to write equation (\ref{dot Us}) in vector form, replacing
the $N \times N$ matrix $U_s(T)$ with the $N^2$ dimensional vector
$\textbf{u}_s$:
\begin{equation}\label{Gmat}
\frac{\dd \textbf{u}_s}{\dd s} =\left[\int_0^T
\frac{\delta\textbf{u}_s}{\delta
\e_s(t)}\frac{\delta{\textbf{u}^{\dag}_s}}{\delta \e_s(t)}\dd
t\right]\nabla \Phi_M(\textbf{u}_s) :=\textmd{F}[\e_s(t)]\nabla
\Phi_M(\textbf{u}_s)
\end{equation}
where the elements of the matrix $F$ are
$$\textmd{F}_{ij,pq} =
-\int_0^T
\big(U_s(T)\mu_s(t)\big)_{ij}\big(\mu_s(t)U^{\dag}_s(T)\big)_{pq}\dd
t.$$
This relation implies that the
variation of the propagator in $\U(N)$ caused by following the
gradient flow in the space of control fields is
Hamiltonian-dependent, with the influence of the Hamiltonian completely
contained in the $N^2$-dimensional positive-semidefinite, symmetric
matrix $\textmd{F}[\e_s(t)]$. We will make further use of this decomposition in the next section.

A second drawback to using gradient flows to search quantum multiobservable control
landscapes is that their convergence rate depends on the properties of the observables
$\Theta_k$ and the initial density matrix $\rho(0)$. This effect is purely kinematic and
does not depend on the Hamiltonian. In the Appendix we explicitly
integrate the kinematic gradient flow (11) in the special case that $\rho(0)$
is a pure state, and show that it follows a convoluted path in $\U(N)$. In general,
the kinematic flow evolves on the interior of a polytope whose
dimension (and the mean path length of the flow trajectory) rises
with rank and eigenvalue nondegeneracy in $\rho(0)$ (Appendix A).
Moreover, it can be shown (Appendix B) that each term in the
multiobservable gradient (\ref{gradmult}) can be expanded on a
``natural" set of basis functions consisting of linear combinations
of matrix elements of the time-evolved dipole operator $\mu(t)$; the
dimension of this basis is:
\begin{equation}\label{gradrank}
D=N^2-(N-n)^2-\sum_{i=1}^r n_i^2 = n(2N-n)-\sum_{i=1}^r n_i^2,
\end{equation}
where $n$ is the rank of $\rho(0)$ and $n_i$ denotes the degeneracy of $i$-th distinct
eigenvalue of $\rho(0)$ (out of the total $r$ such distinct eigenvalues). $D$ also increases
with the rank and eigenvalue nondegeneracy of $\rho(0)$.

Gradient flow control optimization is thus Hamiltonian-dependent and decreases in efficiency for cases where
$\rho(0)$ and the observables $\Theta_k$ are nondegenerate and of high rank.
Despite these drawbacks to gradient flow sampling of
multiobservable control landscapes, more sophisticated
gradient-based algorithms may offer a significant advantage over stochastic search, due to the favorable
critical topology of these landscapes. We consider these algorithms
below.

\section{Unitary matrix flow tracking}\label{unitarytracking}

The symmetric, positive semidefinite matrix
$\textmd{F}\left[\e_s(t)\right]$ in equation (\ref{Gmat}) above
indicates that the convergence time for local gradient-based OCT
algorithms may vary greatly as a function of the Hamiltonian of the
system. Given the decomposition of the gradient into
Hamiltonian-dependent and Hamiltonian-independent parts, the natural
question arises as to whether the Hamiltonian-dependent part can be
suppressed to produce an algorithm whose convergence time will be
dictated by that of the unitary gradient flow (or a suitable kinematic analog), irrespective of the
system Hamiltonian.

In order for the projected flow from $\e_s(t)$ onto $U_s(T)$ to
match the path followed by the gradient flow (\ref{Uflowmult}), the quantity
$\frac{\partial{\e_s(t)}}{\partial s}$ that corresponds to movement
in each step must satisfy a matrix integral equation:
\begin{equation}\label{gendiff}
\frac{\dd U_s(T)}{\dd s} = \int_0^T \frac{\delta U_s(T)}{\delta
\e_s(t)}\frac{\partial{\e_s(t)}}{\partial s}\dd t=\nabla
\Phi_M(U_s(T)).
\end{equation}
In the dipole formulation, this relation becomes
\begin{equation}\label{matint0}
\int_0^T \mu_s(t)\frac{\partial
{\e_s(t)}}{\partial s}\dd t = i\hbar U^{\dag}_s(T)\nabla \Phi_M(U_s(T)),
\end{equation}
where, as in the previous section, $\mu_s(t) \equiv U^{\dag}_s(t)\mu U_s(t).$ This is a Fredholm integral equation of the first kind \cite{Bertero1985}
for $\frac{\partial \e_s(t)}{\partial s}$, given $\e_s(t)$ at $s$ and all $t \in [0,T]$. When
$\Phi_M$ takes the form in equation (\ref{multi}), we have
$$\int_0^T \mu_s(t) \frac{\partial \e_s(t)}{\partial s}\dd t =
i\hbar\sum_{k=1}^m \alpha_k\left[\rho(0),U^{\dag}_s(T)\Theta_k U_s(T)\right].$$  On the basis of
eigenstates, the matrix integral equation (\ref{matint0}) is written
\begin{equation}\label{matint}
\int_0^T \langle i|\mu_s(t)|j\rangle \frac {\partial
\e_s(t)}{\partial s} \dd t = \imath\hbar\langle
i|(U_s(T))^{\dag}\nabla \Phi_M (U_s(T))|j\rangle.
\end{equation}
The relation (\ref{matint}) is underspecified with respect to
$\e_s(t)$, indicating that the integral equation possesses a family of solutions.
To solve it, we must convert it to an explicit initial value problem for $\e_s(t)$
given $\e_0(t)$. We first expand the partial derivative as
\begin{equation}\label{free}
\frac {\partial \e_s(t)}{\partial s} = \sum_{i,j=1}^N
x^{ij}_s ~\langle i|\mu_s(t)|j\rangle+f_s(t)
\end{equation}
on the basis of functions $\langle i|\mu_s(t)|j\rangle $, where the
``free function" $f_s(t)$ contains the additional linearly independent
degrees of freedom. Inserting this expansion into equation
(\ref{matint}) produces
\begin{multline*}\label{insert}
\sum_{p,q=1}^N x^{pq}_s \int_0^T \langle i|\mu_s(t)|j\rangle
\langle p|\mu_s(t)|q\rangle  \dd t \\
= \langle i|\Delta_s|j\rangle-\int_0^T f_s(t)\langle
i|\mu_s(t)|j\rangle\dd t,
\end{multline*} where $\Delta_s=\imath\hbar U^{\dag}_s(T)\nabla \Phi_M(U_s(T))$.
For more general target tracks $Q_s$ in $\U(N)$,  we have $\Delta_s=\frac{\dd Q_s}{\dd s}$.  If we denote the Gramian matrix $\textmd{G}_s$ as
\begin{equation}\label{G matrix}
(\textmd{G}_s)_{ij,pq} = \int_0^T \langle i |\mu_s(t)|j\rangle
\langle p~|\mu_s(t)|q\rangle \dd t,
\end{equation}
we can solve for the coefficients $\vec x= (x^{11}_s,\cdots,x^{1N}_s;\cdots; x^{N1}_s,\cdots,x^{NN}_s)^T$ as
$$\vec x= \textmd{G}_s^{-1}\left(\frac{\dd Q_s}{\dd s}-\int_0^T\mu_s(t')f_s(t')\dd t'
\right),$$ provided that $\textmd{G}_s$ is invertible. We then
obtain the following initial value problem for $\e_s(t)$, the algorithmic
evolution of the control field along the track:
\begin{multline}\label{utrack}
\frac{\partial \e_s(t)}{\partial s} =  f_s(t) +\\
+ v\left(\frac{\dd Q_s}{\dd s} - \int_0^T \mu_s(t' ) f_s(t' ) \dd t'
\right)^T\textmd{G}_s^{-1}v({\mu_s(t)}),
\end{multline}
where the operator $v(\cdot)$ vectorizes its matrix argument.
Each ``free function" $f_s$ corresponds to a unique algorithmic step in $\e(\cdot)$;
modulating this function allows for systematic exploration of the
set of functions $\e_s(t)$ that are compatible with the gradient
step on $\U(N)$ \cite{Dominy2008}.

Solving this set of $N^4$ scalar differential equations requires
that the $N^2 \times N^2$ matrix $\textmd{G}_s$ defined by (\ref{G
matrix}) is invertible, which is equivalent to the claim that the
map from between control fields and unitary propagators is locally
surjective in a sufficiently small neighborhood of $U_s$. Control
fields at which $\textmd{G}_s$ is singular correspond to so-called
\textit{singular extremal} solutions to the control problem, in order to
contrast them from \textit{regular extremals} \cite{Bonnard2003}. Even if
$\textmd{G}_s$ is invertible, it is possible that it is nearly
singular, resulting in large numerical errors during the solution to
the differential equation. It is convenient to assess the nearness
to singularity of $\textmd{G}_s$ by means of its condition number,
namely the ratio of its largest singular value to its smallest
singular value.

Although the gradient function $\frac{\delta \Phi_M}{\delta\e_s(t)}$
is always locally the direction of fastest increase in the objective
function at $\e_s(t)$, the path $\e_s(t)$ (parameterized by $s$) derived from
following this gradient has no universal (Hamiltonian-independent)
global geometry, since $\Phi_M$ is not explicitly a function of
$\e_s(t)$. It is known \cite{Mike2006a} that this path will not
encounter any traps during the search, but
beyond this, the geometry can be expected to be rugged.
Unlike the dynamical gradient flow (\ref{Egrad}), the algorithmic flow
(\ref{utrack}) follows the gradient flow on $\U(N)$.
This flow respects the geometric formulation of the optimal control
objective function in terms of $U_s(T)$ rather directly in terms of
$\e_s(t)$. The functions $\mu_s(t)$ contain all relevant
information about the quantum dynamics, whereas the functions
$\frac{\dd Q_s}{\dd s}$ contain complete information
about the geometry of the kinematic search space. The $N^2$ functions
$\langle i|\mu_s(t)|j\rangle$, $1\leq i,j\leq N$, are calculated
during the evaluation of $\frac{\delta \Phi_M}{\delta\e_s(t)}$; hence,
the computational overhead incurred by following this flow
corresponds to that needed to compute the $N^4$ elements of
$\textmd{G}_s$ and invert the matrix, at each algorithmic time step.

A multitude of other flows $\frac{\dd Q_s}{\dd s}$ could be substituted in the RHS of
equation (\ref{matint}). Since we are interested in global
optimality, we should choose a flow that follows the shortest
possible path from the initial condition to a unitary matrix that
maximizes the observable expectation value. It can be shown
\cite{Mike2006a} that a continuous manifold of unitary matrices $W$
maximizes $\Phi_{(M)}$. These unitary matrices can be determined
numerically by standard optimization algorithms on the domain of
unitary propagators \cite{Brockett1991}. The shortest length path in
$\U(N)$ between the initial guess $U_0$ and an optimal $W$ is then
the geodesic path, which can be parameterized as
\begin{equation}\label{geod}
Q_s=U_0\exp(\imath As)
\end{equation}
with $A = -\imath\log(W^{\dag}U_0)$, where $\log$ denotes the complex
matrix logarithm with eigenvalues chosen to lie on the principal
branch $-\pi<\theta<\pi$. Thus, if we set $\Delta_s = A$, the
tracking algorithm will attempt to follow the geodesic path
\footnote{In the case that the control system evolves on a subgroup
of $\U(N)$, e.g. $\mathcal{S}\mathcal{U}(N)$, the geodesic on that subgroup can be tracked
instead.}.

Due to the nonlinearity of the integral equation (\ref{gendiff}), errors
in tracking will inevitably occur, increasing the length of the
search trajectory beyond that of the minimal geodesic path. These tracking errors
may be ameliorated through the introduction of stabilization terms in equation (\ref{utrack}) (see Section \ref{err}) or higher order functional derivatives in equation (\ref{matint0}).

\section{Multiobservable homotopy tracking control}\label{orthog}

As a methodology for multiobservable control, unitary matrix tracking has an advantage
over gradient search in that it can directly follow an optimal path in the space of unitary
propagators, assuming the map $\varepsilon(t) \mapsto U(T)$ is surjective and the
first-order formulation of the tracking equation (\ref{utrack}) is sufficiently accurate.
However, it cannot be implemented experimentally
without expensive tomography measurements, and carries a
computational overhead that scales exponentially with system size.

We now consider a related, \textit{experimentally implementable} tracking
control algorithm --- multiobservable homotopy tracking control
(MOTC) --- that seeks to drive the expectation values of $m$ observable
operators $\Theta_1,...,\Theta_m$ along a predetermined path
$\textbf{w}_s$ of $\vec{\Phi}$ to desired target values. These
trajectories may correspond to expectation value paths corresponding
to the kinematic gradient flow (\ref{Uflowmult}), the geodesic
(\ref{geod}), or any other path through multiobservable space. In
particular, unlike the gradient flows (\ref{multi}), MOTC may be
used to successively drive the expectation values of individual
observables to their maxima, while constraining the others to fixed
values.

The observables measured at each step can be assumed to be linearly
independent without loss of generality. Denote the $m$ scalar
functions of algorithmic time (expectation value paths for each
observable) by $\Phi_s^1,\ldots, \Phi_s^m$. Then, the vector
Fredholm integral equation for $\frac{\partial \e_s(t)}{\partial s}$
analogous to equation (\ref{gendiff}) is given by
\begin{equation}\label{gendiff2}
\frac{\dd \Phi_s^i}{\dd s} = \int_0^T \frac{\delta \Phi_s^i}{\delta
\e_s(t)}\frac{\partial{\e_s(t)}}{\partial s}\dd t=\frac{\dd \textbf{w}_s^i}{\dd s}, \quad 1 \leq i \leq m.
\end{equation}
where $\Phi_s^i=
\tr(U_s(T)\rho(0)U_s^\dag(T)\Theta_i)$. As in section \ref{unitarytracking}, to solve for the algorithmic flow
$\frac{\partial \e_s(t)}{\partial s}$ that tracks this path, it is necessary to expand it on a basis
of independent functions. In this case, we make this expansion on the
basis of the independent observable expectation value gradients:
\begin{equation}\label{expansion}
\frac{\partial \e_s(t)}{\partial
s} = \sum_{j=1}^m x^{j}_s \frac{\delta \Phi_s^j}{\delta
\e_s(t)}+f_s(t),
\end{equation}
where the free function is similar to that in equation (\ref{free}).
Inserting the expansion (\ref{expansion}) into the resulting generalized differential equation, we have
$$\sum_{j=1}^m x^{j}_s \int_0^T\frac{\delta \Phi_s^i}{\delta \e_s(t)}~
\frac{\delta \Phi_s^j}{\delta \e_s(t)}\dd t= \frac{\dd
\textbf{w}^{i}_s}{\dd s}-\int_0^T \frac{\delta \Phi_s^i}{\delta
\e_s(t)}f_s(t)\dd t.$$ Defining the $m$-dimensional vector
$\textbf{a}_s(t)=(a_s^1(t),\cdots,a_s^m(t))^T$ by
\begin{eqnarray}\label{adef}
a_s^i(t)&=&\frac{\delta \Phi_s^i}{\delta \e_s(t)}=-\frac{\imath}{\hbar}\tr\left(U^{\dag}_s(T)\mu_s(t)\nabla\Phi^i(U_s(T))\right)\\
&=&-\frac{\imath}{\hbar}\tr\left(\rho(0) \big[U^{\dag}_s(T)\Theta_iU_s(T),\mu_s(t)\big]\right),
\end{eqnarray}
and the MOTC Gramian matrix $\Gamma$ as
$$(\Gamma_s)_{ij} = \int_0^T a_s^i(t)a_s^j(t)\dd t,$$
we can then solve for the expansion coefficients $\vec
x=(x^1_s,\cdots,x^m_s)^T$ as
$$\vec x= \Gamma_s^{-1}\left[\frac{\dd
\textbf{w}}{\dd s}-\int_0^T\textbf{a}_s(t')f_s(t')\dd t'
\right],$$
provided that $\Gamma_s$ is invertible. Returning to the
original expansion (\ref{expansion}), we obtain the following
nonsingular algebraic differential equation for the algorithmic flow of the control field:
\begin{equation}\label{vectrack}
\frac {\partial \e_s(t)}{\partial s} = f_s(t) + \left[\frac{\dd
\textbf{w}_s}{\dd s}-\int_0^T\textbf{a}_s(t')f_s(t')\dd t'
\right]^T\Gamma_s^{-1}\textbf{a}_s(t).
\end{equation}

In the special case where only one observable $\Theta$ is measured
at each algorithmic step, this equation reduces to:
\begin{equation}\label{scalarflow}
\frac {\partial \e_s(t)}{\partial s} = f_s(t)+
\frac{a_s(t)}{\gamma_s}\left({\frac{\dd P_s}{\dd s} -\int_0^T a_s(t'
)f_s(t' ) \dd t' }\right),
\end{equation}
where $P_s$ is the desired track for $\langle \Theta(T) \rangle$,
and $\gamma_s =\int_0^T a^2_s(t)\dd t$.
We note that unitary matrix tracking can also be expressed as a special case of equation
(\ref{vectrack}), if the functions $\Phi^{i}$ are taken to be the matrix elements of $U_s(T)$, i.e.,
$$\Phi^{(j-1)N+k}(U_s(T))=\langle j|U_s(T)|k\rangle,~~~j,k=1,\cdots,N.$$
Then, according to equation (\ref{adef}), the $N^2$-dimensional complex vector
$\textbf{a}_s(t)=v(\mu_s(t))$ is the vector form of the matrix
$\mu_s(t)$:
$$a_s^{(j-1)N+k}(t)=\frac{1}{\imath\hbar}\langle
j|\mu_s(t)|k\rangle, ~~~j,k=1,\cdots,N.$$ With this choice, the $N^2
\times N^2$ Gramian matrix $\Gamma$ is identical to ${\rm G}_s$ as defined by (\ref{G matrix}). Indeed,
with $\Phi^i$ and $a^i$ suitably defined, the MOTC tracking differential equation (\ref{vectrack}) provides
a unified framework for continuation approaches \cite{Hillermeier2001} to generic quantum multiobjective control problems.

One disadvantage of multiobservable tracking, compared to
local gradient optimization of objective function $\Phi_M$, is that
it encounters singular control fields more frequently. In this case, singularities correspond to the
situation that variation of control fields near the control under
consideration is not sufficient to produce arbitrary variation of
$\vec \Phi(U(T))$ driven by this control. Nonetheless, the likelihood of the matrix
$\Gamma_s$ being ill-conditioned - even at singular \textit{extremal} control
fields $\e(t)$, where $\textmd{G}$ is singular - diminishes
rapidly with $N^2-m$ (Section \ref{numerical}).

Auxiliary penalties on $\varepsilon_s(t)$, such as practical
experimental constraints on the total field fluence, act to decrease
the degeneracy in the solutions to the above system of equations for
$\frac {\partial \e_s(t)}{\partial s}$.
It can be shown \cite{Rothman2005} that the choice:
\begin{equation}\label{minfluence}
f_s(t) = - \frac{1}{\eta_s} \e_s(t) w(t),
\end{equation}
for the free function in the tracking differential equations, where
$w(t)>0$ is an arbitrary weight function and the $\eta_s$ term (typically
constant) controls numerical instabilities, will determine the $\frac
{\partial \e_s(t)}{\partial s}$ at each algorithmic time step $s$
that minimizes fluence.

\section{Error correction}\label{err}

Errors can occur when tracking paths on $\U(N)$ or subspaces of its
homogeneous spaces for two reasons. First, the algorithmic steps on
these spaces will be first-order approximations to the true
increments ($Q_{s_{k+1}}-Q_{s_{k}}$ or
$\textbf{w}_{s_{k+1}}-\textbf{w}_{s_{k}}$) due to discretization
error; this error will increase as a function of the curvature of
the flow trajectory $Q_s$ or $\textbf{w}_s$ at algorithmic time
$s_k$. Second, the tracking integral equations are formulated in
terms of only the first-order functional derivatives $\frac {\delta
U_s(T)}{\delta \e_s(t)}$ or $\frac {\delta \vec\Phi}{\delta
\e_s(t)}$; the truncation error will depend on the magnitude of
higher order functional derivatives.

In numerical simulations, error-correction methods can be applied to
account for these deviations from the track of interest. Generally,
we choose an error correction term $\mathbf{c}_s$ that reflects the
difference between the current value of the tracking vector and
its target value, such that the tracking differential equation (\ref{vectrack})
becomes
\begin{equation}\label{motcerr}
\frac {\partial \e_s(t)}{\partial s}= f_s(t) +
\left[\mathbf{c}_s+\frac{\dd \textbf{w}_s}{\dd s}-
\int_0^T\textbf{a}_s(t')f_s(t')\dd
t'\right]^T\Gamma_s^{-1}\textbf{a}_s(t).
\end{equation}

For unitary matrix tracking, one can follow the (minimal-length) geodesic from the real point
$U_{s_k}(T)$ to the track point $Q_{s_k}$ \cite{Dominy2008}. This
correction can be implemented through the choice
$$\mathbf{c}_{s_k}=v\left(-\frac{\imath}{s_{k+1}-s_k}\log\left[Q^{\dag}_{s_k}U_{s_k}(T)\right]\right)$$
in the discretized version of (\ref{motcerr}).

For general multiobservable expectation value tracking, the vector
space within which $\vec\Phi_s$ resides is not a Lie group, but
rather a subspace of a homogeneous space, and consequently it is not
as straightforward to apply error correction algorithms that exploit
the curved geometry of the manifold \footnote{A Riemannian metric on the
space of density matrices may be defined \cite{Uhlmann1976}, but its application
to multiobservable error correction when $m<N^2-1$ is nontrivial.}. In this paper,
we take as the correction term a simple scalar multiple of the difference between the
actual value and the target track, i.e.,
$$\mathbf{c}_s=\beta(\mathbf{w}_s-\vec\Phi_s),~~~\beta>0.$$

Note that these methods can in principle also be implemented in an
experimental setting. The use of Runge-Kutta integrators (RK, see below) to solve the MOTC
differential equation (\ref{motcerr}) with error correction often
decreases tracking errors considerably, especially for problems
involving integration over long $s$ intervals. RK integration with
error correction is the method of choice for numerical
implementations of MOTC, although it is more difficult
to apply experimentally.

\section{Numerical implementation}\label{methods}

For illustrations of multiple observable tracking, a set of $m<N$ commuting operators
$\Theta_1,\cdots, \Theta_m$ was employed. $\Theta_1$ was a randomly chosen nondegenerate diagonal matrix,
and $\Theta_2,\cdots, \Theta_m$ were sequential pure state projection operators in the canonical basis. The $m$-dimensional vector ${\vec\Phi}_s$ was constructed by
taking the trace of the product of each of these observable operators with the time-evolved density matrix.
Numerical solution of the tracking differential equations
(\ref{vectrack}, \ref{scalarflow}) was carried out as follows. The
electric field $\e_s(t)$ was stored as a $p\times q$ matrix, where
$p$ and $q$ are the number of discretization steps of the
algorithmic time parameter $s$ and the dynamical time $t$,
respectively (i.e., for each algorithmic step $s_k$, the field was
represented as a $q$-vector for the purpose of computations).
Starting from an initial guess $\e_{s_0}(t)$ for the control field,
the \sd equation was integrated over the interval $~[0,T~]$ by
propagating over each time step, producing the
local propagator $$U(t_{j+1},t_j) = \exp~[-\imath
H_{s_k}(t_j)T/(q-1)~].$$ The propagation toolkit \cite{Yip2003}
was used for this purpose. Local propagators were precalculated via diagonalization
of the Hamiltonian matrix (at a cost of $N^3$), exponentiation of
the diagonal elements, and left/right multiplication of the
resulting matrix by the matrix of eigenvectors/transpose of the
matrix of eigenvectors. This approach is generally faster than
computing the matrix exponential directly. Alternatively, a Runge-Kutta integrator \cite{Press2002}
can be employed for time propagation.

The time propagators $$U(t_j,0)=U(t_j,t_{j-1})\cdots U(t_1,0)$$
computed in step 1 were then used to calculate the time-evolved
dipole operators $\mu(t_j)=U^{\dag}(t_j,0)\mu U(t_j,0)$, which can
be represented as a $q$-dimensional vector of $N\times N$ Hermitian
matrices. The $\Gamma_{s_k}$ and vector $\textbf{a}_{s_k}$ were then
computed by time integration of the dipole functions with an
appropriate choice of function $f_s(t)$ described above; in the present work, $f_s$ was
set either to 0 or the expression in (\ref{minfluence}).  For
tracking of unitary flows - either the kinematic gradient flow
(11) or the geodesic flow (\ref{geod}) - the next point
$Q_{s_{k}}$ on the target unitary track was calculated numerically
through $Q_{s_k}=Q_{s_{k-1}}\exp(-\imath\Delta_{s_k} \dd s)$, using
a matrix exponential routine.

Next, the control field $\e_{s_k}(t)$ was updated to
$\e_{s_{k+1}}(t)$. This step required inversion of the $N^2
\times N^2$ matrix $\textmd{G}_{s_k}$ or $m \times m$ matrix
$\Gamma_{s_k}$, which was carried out using LU decomposition. The
quantities $\Gamma^{-1}_{s_k}$, $\textbf{a}_{s_k}$ and $\textbf{c}_{s_k}$
were used to compute the $q$-dimensional vector $\frac{\partial
\e_{s_{k}}(t)}{\partial s}$. One of two approaches was used to
update the field: (i) a simple linear propagation scheme, i.e.,
$$\e_{s_{k+1}}(t) = \e_{s_{k}}(t) + (s_{k+1}-s_k) \frac{\partial
\e_{s}(t)}{\partial s}\Big|_{s=s_k},$$ or (ii) a fourth- or
fifth-order Runge-Kutta
integrator. The updated control field $\e_{s_{k+1}}(t)$ was then again used to the propagate
the \sd equation.

Adaptive control of step size was used to accelerate convergence for
both the gradient flow and MOTC algorithms. For the gradient flow,
the minimum of the objective function $\Phi$ or $\Phi_M$ along the
direction of the gradient at step $s$ was located by first
bracketing the minimum in that direction by parabolic extrapolation,
followed by application of Brent's method for inverse parabolic
interpolation to pinpoint the minimum. For both MOTC and the
gradient flow, the Fehlberg embedded Runge-Kutta integrator (ASRK5)
with Cash-Karp parameters \cite{Press2002} was used, which embeds a
fourth-order integrator within a fifth order integrator, and
compares the differences between the fourth- and fifth-order
estimates to assess truncation error and adjust $\Delta s$. To
compare efficiency of MOTC and gradient control algorithms, the
minimum step size and error tolerance level in ASRK5 were set to the
largest values that permitted stable integration of at least one of
the algorithms.

\section{Examples}\label{numerical}

\begin{figure*}
\centerline{
~~~~~~~~~~~~~~~~~\includegraphics[width=7.7in,height=5in]{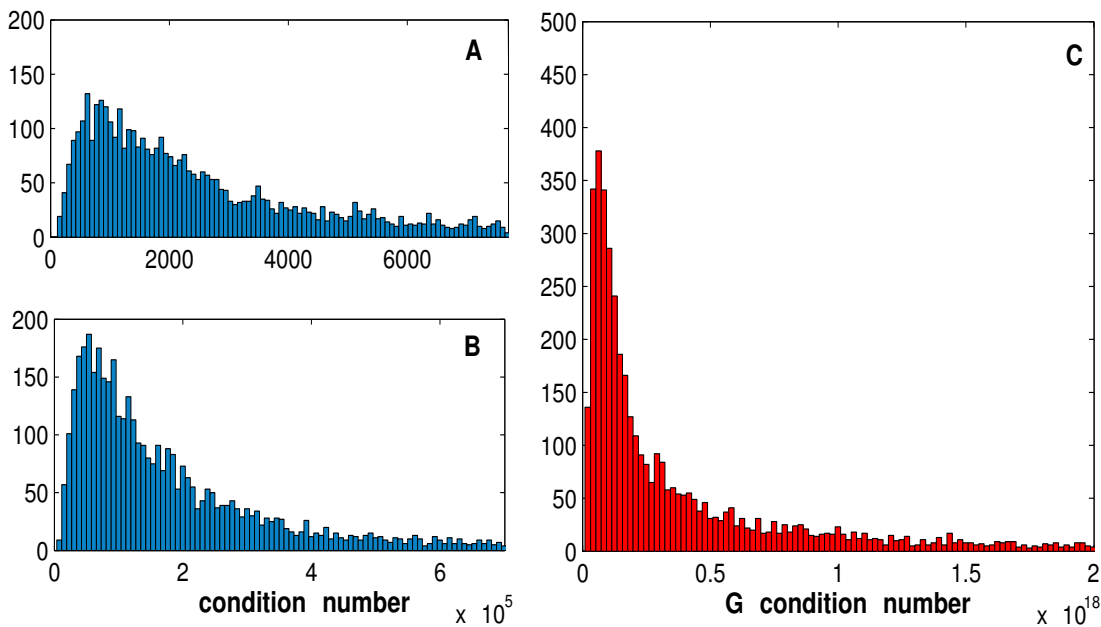}
} \caption{Gramian matrix condition number distributions for the
11-level system in section \ref{numerical}. The amplitudes and
phases of the modes of the control field $\varepsilon(t)$ were
sampled randomly from the uniform distributions $(0,1]$ and
$(0,2\pi]$, respectively, with the mode frequencies tuned to the
transition frequencies of the system. \textbf{(A)} MOTC $\Gamma$
matrix with 10 observables and $\rho(0)$ being full-rank and
nondegenerate (thermal); \textbf{(B)} MOTC $\Gamma$ matrix with 10
observables and $\rho(0)$ being a pure state; \textbf{(C)} Unitary
tracking $\mathrm{G}$ matrix. }\label{gramfig1}
\end{figure*}

As an illustrative example of multiobservable control, we examine the problem of targeting the vector
of $m$ observable expectation values associated with the unitary
propagator $W$ to which the kinematic gradient of a single observable converges. We seek to
track the multiobservable path $\textbf{w}_s$ of expectation values that corresponds to the geodesic between
$U_0$ and $W$, i.e. $w_s^k = \tr\left\{\rho(0) \exp(-\imath As)U^{\dag}_0 \Theta_k U_0\exp(\imath As)\right\}$, with $A = -\imath\log(W^{\dag}U_0)$,$~1 \leq k \leq m$ for various numbers of observables $m$. Although this is a simple incarnation of multiobservable quantum control, it is well-suited for illustration
of its basic principles, since it deals with a universally
applicable control objective, and allows for a systematic study of
the effects of the imposition of additional observable objectives on
the search dynamics and optimal control fields.

The examples below employ an 11-dimensional Hamiltonian of the form (\ref{ham}), with
\begin{eqnarray}\label{system example}
H_0 &=& diag~\{0.1,0.2,\cdots,1.1\},\\
\mu_{ij}&=&\left\{
\begin{array}{cl}
1,& i=j;\\
0.15,&|i-j|=1;\\
0.08,&|i-j|=2;\\
0, &\text{otherwise}.
\end{array}\right.
\end{eqnarray} We first assess the properties of the Gramian matrices $\textmd{G}$ and
$\textmd{Gamma}$ for such systems.

\subsection{MOTC Gramian matrix estimation}

An important consideration in any deterministic multiobservable
quantum control optimization is the singularity of control fields
applied to the quantum system, where the singularity of the mapping
$\e(t)\mapsto U(T)$ (assessed through the Gramian $\textmd{G}$)
must be distinguished from that of the mapping $\e(t)\mapsto \vec\Phi(U(T))$ (assessed through
the Gramian $\Gamma$). Nonsingularity of the latter corresponds to the ability to move,
through an infinitesimal change of the control field $\varepsilon_s(t) \rightarrow \varepsilon_{s+\dd s}(t)$, between two infinitesimally close vectors of multiple observable expectation
value vectors, $\vec \Phi \rightarrow \vec \Phi+ \dd \vec \Phi$. The
Gramian \textmd{G} depends only on $H_0$, $\mu$ and $T$, whereas $\Gamma$
additionally depends on the eigenvalue spectra of $\rho(0)$ and
$\Theta_k$. Experimentally or numerically, an ill-conditioned
Gramian matrix will result in large tracking errors for
$\varepsilon_s(t)$.

The requirements for a control field to be regular for the mapping
$\e(t)\mapsto U(T)$ are, in general, more stringent than those
for the mapping $\e(t)\mapsto \vec\Phi(U(T))$, since multiobservable
control requires control of a subset of the parameters of $U(T)$.
However, if $\rho(0)$ is rank-deficient, the multiobservable Gramian
matrix $\Gamma$ can become more ill-conditioned than $\textmd{G}$,
since certain paths $U_s(T)$ cannot be accessed. Fig. \ref{gramfig1}
compares the condition number distributions for $\textmd{G}$ and
$\Gamma$ for various $\rho(0)$, where the number of observables
$m=10$, for randomly sampled fields $\varepsilon(t)$ of the form
\begin{equation}\label{randfield}
\varepsilon(t) = \sum_{i=1}^N \sum_{j=i+1}^N A_{ij}\sin\left(\omega_{ij}t+\phi_{ij}\right), \quad 0\leq t \leq T,
\end{equation}
where $\omega_{ij}=|E_i-E_j|$ denote the transition frequencies
between energy levels $E_i,~E_j$ of $H_0$, $\phi_{ij}$ denotes a
phase sampled uniformly within the range $(0,2\pi]$, and $A_{ij}$
denotes a mode amplitude sampled uniformly within the range $(0,1]$.
The final time $T$ was chosen to be sufficiently large to achieve
full controllability over $\U(N)$ at $t=T$ \cite{Raj2007}.

The absolute magnitudes of the condition numbers for a given
$\rho(0)$ spectrum depends on the Hamiltonian, with random, dense
$H_0$, $\mu$ matrices generally exhibiting more well-conditioned
Gramian matrix distributions.
Typically, a Gramian condition number $C >
10^8$ results in large numerical errors upon inversion, and would be
expected to compromise the accuracy of tracking (due to the
sparseness of control field increments $\delta \varepsilon(t)$
that are capable of driving the system to the corresponding state).

For small numbers of observables $m$, rank-deficiency in $\rho(0)$
does not shift the condition number distribution for $\mathrm{G}$
toward substantially higher values, compared to that of $\Gamma$,
since the number of parameters of $\rho(T)$ that must be controlled
is small. For example, the pure and thermal mixed states in Fig.
\ref{gramfig1} both have well-conditioned Gramians $\Gamma$, which
permit accurate multiobservable tracking (see below).

For a quantum ensemble in thermal equilibrium with a bath at
temperature $T_e$, the eigenvalues are determined by the Boltzmann
distribution,  i.e.,
$$\lambda_i = \frac{\exp(-E_i/kT_e)}{\sum_{i=1}^N\exp(-E_i/kT_e)},\quad i=1,\ldots,N,$$ and  MOTC can be
carried out without the additional overhead of density matrix
estimation.  Through its effect on the eigenvalues of $\rho(0)$, the
spectrum of $H_0$ (in particular, the energy-level spacings) plays
an important additional role in determining the Gramian matrix
condition number distribution for thermally prepared states.

The regularity of control fields generally becomes harder to achieve for
larger Hilbert space dimension. For example, for Hamiltonians of the
form (\ref{ham}), the mean $\mathrm{G}$ condition number rose from
$\mathcal{O}(10^{18})$ for $N=11$ to $\mathcal{O}(10^{19})$ for
$N=19$, and the $\Gamma$ condition number for full rank $\rho$ rose
from $\mathcal{O}(10^5)$ to $\mathcal{O}(10^6)$ for tracking of one
complete measurement basis ($m=10$, $m=18$, respectively).

The distributions above provide rough estimates for probabilities of
encountering singularities during the implementation of the
different forms of MOTC examined in the following sections. However,
the actual changes in $\Gamma$ that occur along an optimization
trajectory will not be identical to those obtained by random
sampling from these distributions, since the frequencies $\omega$ of
successive control fields in MOTC are not independent random
variables \footnote{The maximum change in $\Gamma$ (and hence its condition number $C$) that
can occur along an optimization trajectory is bounded due to the fact that the norm of the
gradient $\frac{\delta \Phi_k}{\delta \varepsilon(t)}$ is
bounded by the norm of the dipole operator $||\mu||$ \cite{HoRab2007a}.}
As such, the fields $\varepsilon_s(t)$ sampled during MOTC tracking will typically be
more regular than those in the distributions displayed above, although the trends will be similar.

\subsection{Multiobservable tracking}

\begin{figure}
\centerline{
~~~~~~~~~~~~~\includegraphics[width=7.5in,height=5.5in]{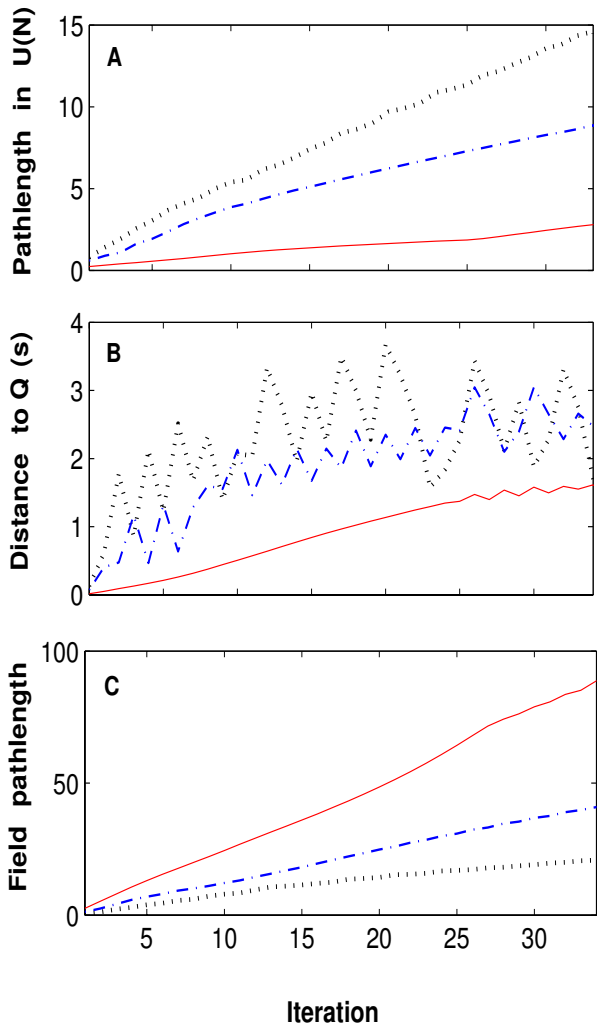}
} \caption{Comparison of MOTC of 2 (dotted), 4 (dashed) and 10
(solid) observables from a single measurement basis, for 11-level
system (\ref{system example}). $\rho(0)$ was of rank 7 with
nondegenerate eigenvalues. The target was a unitary propagator that
maximized the expectation value of the first observable. \textbf{A)}
Total distance (Frobenius norm) in $\U(N)$ traversed by the dynamical propagator $U(T)$
versus algorithmic step; \textbf{B)} Distance between target $Q_s$ and actual $U_s(T)$
tracks as function of algorithmic step; \textbf{C)} Total distance (Euclidean norm) in $L^2(\R)$ between the
current field $\varepsilon_s(\cdot)$ and the original field
$\varepsilon_0(\cdot)$ along the optimization trajectory.
}\label{onebasis1}
\end{figure}

\begin{figure}
\centerline{
\includegraphics[width=3.5in,height=4.5in]{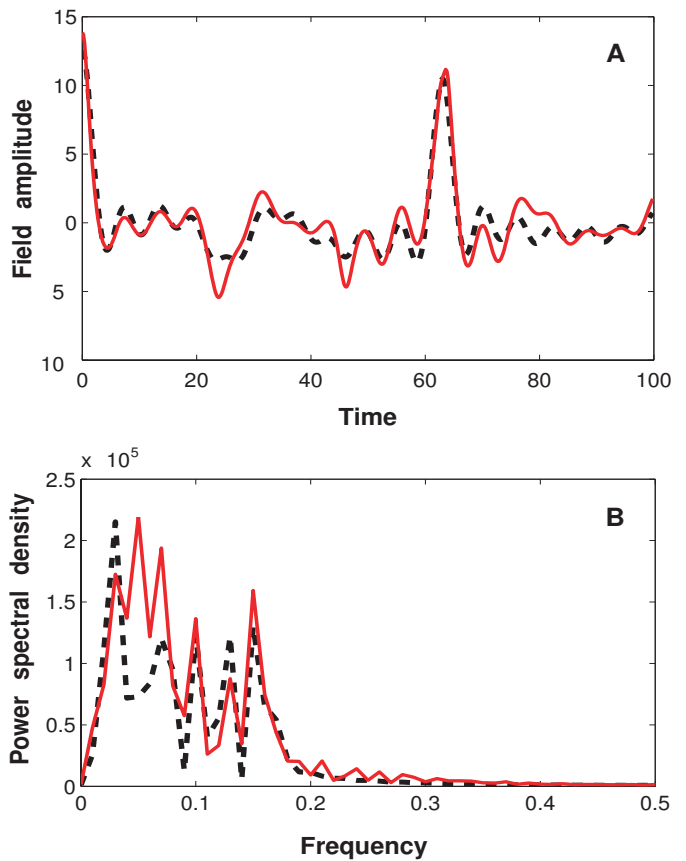}
}
\caption{Comparison of optimal fields obtained via $m=2$ (dashed) and $m=10$ (solid)
 MOTC search algorithms that maximize a single observable $\Theta$. $\rho(0)$ was
  a thermal mixed state. \textbf{A)} Optimal control fields for 2 and 10 observable tracking;
  \textbf{B)} Fourier power spectra of the optimal control fields. Note the increase in high
  frequency modes due to the imposition of additional observable objectives.
  $\varepsilon_0(t)$ was of the form (\ref{randfield}).
}\label{fieldcomp}
\end{figure}

According to equations (\ref{grad1}) and (\ref{gradrank}), the efficiency of gradient flow-based
observable maximization decreases with higher rank and nondegeneracy in $\rho(0)$. In these cases, the gradient flow
typically follows a longer path in both $\U(N)$ and
$\varepsilon(t)$. On the other hand,  the unitary path can be constrained more effectively
by MOTC in such cases, with the most stringent control possible for
full-rank, nondegenerate $\rho(0)$. Fig. \ref{onebasis1} compares
the pathlengths in $\U(N)$ of MOTC optimization trajectories in such a case with 2, 4 and 10
observable expectation values tracked along the geodesic to an optimal $W$ precomputed by following the
kinematic gradient flow $\frac{\dd V_s}{\dd s} = [\Theta_1,V_s\rho(0) V_s^{\dag}]V_s$ \footnote{Note that
although the tracked unitary path $Q_s$ is significantly shorter
than that followed by the dynamical gradient, still shorter paths
could be identified by minimizing the distance of $W$ to $U_0$ while
constraining $\langle \Theta \rangle$ to remain at its maximal
value.}. The Hamiltonian (\ref{system
example}) was employed for these simulations. Across all cases, the Gramian $\Gamma_s$ was well-conditioned
at each step of MOTC optimization, as predicted by the $\Gamma$ condition number distributions above.

As can be seen from the Figure, the pathlength in $\U(N)$ decreases
progressively with increasing $m$; this pathlength is in all cases smaller
than that of the gradient flow (data not shown). For small $m$,  there are significant
stochastic fluctuations in the unitary stepsize per iteration, which
are smoothed out for larger $m$. By contrast, the Euclidean
pathlength traversed in the space of control fields $\varepsilon(\cdot)$
(assessed in terms of the Euclidean distance
$|\varepsilon_{s_1}(\cdot) - \varepsilon_{s_2}(\cdot)|^2 \equiv
\sum_i |\varepsilon_{s_1}(t_i) - \varepsilon_{s_2}(t_i)|^2$, where $s_1,~s_2$ label successive
points along the optimization trajectory) increases systematically with $m$,
with the smallest change in $\varepsilon(\cdot)$ along the optimization trajectory
occurring for the gradient and the greatest change occurring for $m=10$. Almost
universally, imposing additional observable tracks increases the
field pathlength and distance between $\varepsilon_0(\cdot)$ and
$\varepsilon_s(\cdot)$. Since the MOTC field update (\ref{vectrack}) is based on
projections the multiobservable gradient (\ref{gradmult}), each
additional observable tracked causes the optimization trajectory to
deviate further from that followed by the gradient.

Fig. \ref{fieldcomp} depicts the optimal control fields identified by
$m=2$ and $m=10$ MOTC algorithms. Note that, even in cases such as this where
the imposition of multiple observable expectation value constraints drives the
system to a final unitary propagator $W$ that is closer to $U_0$, the optimal
fields are invariably more complex, with the Fourier spectra displaying higher
frequency modes for additional observables. The dissimilarity of
these fields from the initial guess indicates that optimal fields
for multiobservable quantum control cannot be identified by
heuristic reasoning based solely on, for example, the spectrum of
the Hamiltonian.

One disadvantage of tracking-based optimization of observable expectation values, compared
to steepest ascent search, is that more precise measurements of the
gradient are required to remain on the desired track. Tracking paths
for additional observables, where the auxiliary tracks are chosen so
as to constrain the unitary propagator to more a uniform path, can
confer additional stability and robustness to observable
tracking-based optimization.

\begin{figure}
\centerline{
\includegraphics[width=3.5in,height=2.7in]{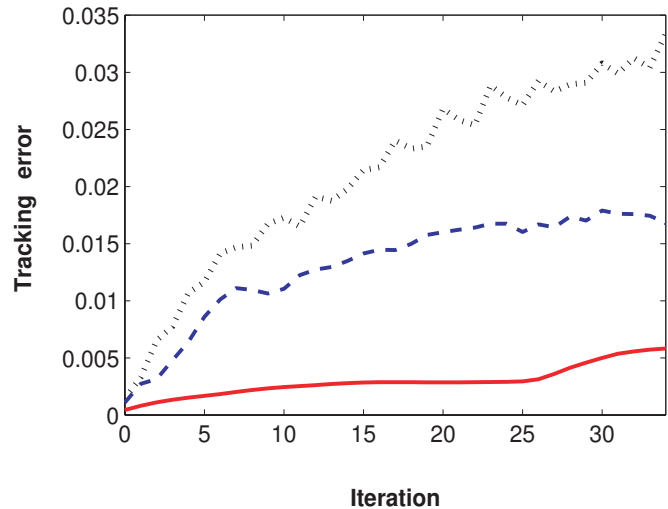}
} \caption{Observable tracking errors for the control problems
displayed in Fig. \ref{onebasis1} over the optimization trajectory.
Dotted = 2 observables; dashed = 4 observables; solid = 10
observables.
}\label{error}
\end{figure}

Errors in tracking that occur due to breakdown of the first-order
MOTC approximation may be stabilized by the use of additional
observables because it is less likely that the first-order
approximation will break down simultaneously for all observables.
Since more auxiliary observables can be tracked when $\rho(0)$ has
more independent parameters, this stabilization method is most effective when
$\rho(0)$ is full-rank. Thus, in multiobservable tracking, a greater
rank of $\rho(0)$ can increase the the maximal stability of the
algorithm, as well as the diversity of possible multiobservable
expectation value targets and the freedom to follow arbitrary paths
toward those targets. Fig. \ref{error} compares the tracking errors for the
MOTC variants depicted in Fig. \ref{onebasis1}.

\begin{figure}
\centerline{
\includegraphics[width=3.3in,height=3.3in]{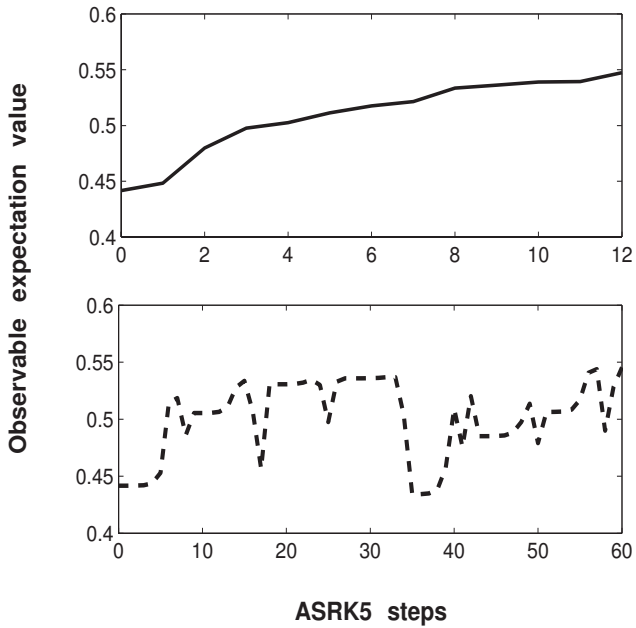}
}
\caption{
Algorithmic efficiency of observable control algorithms assessed through variation of step size.
\textbf{Top}: adaptive stepsize Runge-Kutta (ASRK5) MOTC integration (10 observables);
\textbf{Bottom}: ASRK5 gradient flow integration, both for the observable maximization problem in Fig. \ref{onebasis1}. Expectation value refers to the first observable.
}\label{onebasis1b}
\end{figure}

The relative efficiency of MOTC and gradient flow
observable optimization can be assessed by employing ODE integrators
with adaptive stepsize. For this purpose, the Fehlberg embedded
Runge-Kutta (ASRK5) method was used.  Although it is somewhat more difficult
to implement RK integrators experimentally, they provide a measure of the maximum
stepsize that can be taken along a given optimization trajectory
without incurring errors above a specified tolerance. Fig.
\ref{onebasis1b} compares the number of ASRK5 iterations needed to
solve the above maximization problem by integrating a) the gradient
flow ODE (\ref{Egrad}) and b) the MOTC flow (\ref{vectrack}) with
$\textbf{w}_s$ set to the 10-observable expectation value track corresponding
to the geodesic between $U_0$ and $W$. Another adaptive step-size algorithm for improving optimization efficiency,
a line search (Section \ref{methods}), was also used in the case of the gradient (data not
shown; whereas either ODE can be integrated via ASRK5, only the
gradient flow can be simply implemented via a line search routine).
The substantial decrease in the number of required iterations in the
case of tracking algorithms, indicative of enhanced global
optimality of the path they follow in $L^2(\mathbb{R})$, is
displayed in the Figure.

\section{Discussion}\label{discussion}

We have presented a class of deterministic algorithms for the
optimal control of multiple quantum observables, based on tracking
multiobservable expectation value paths to desired endpoints. These
algorithms leverage the previously reported simple critical topology
of quantum control landscapes \cite{Raj2007}, which indicates that
gradient-based algorithms will rarely encounter singularities. An
additional feature of multiobservable control landscapes that
governs the ability of local search algorithms to follow arbitrary
paths in multiobservable space, the MOTC Gramian matrix, has been
identified and its properties investigated numerically for selected
problems. Error correction methods have been described that should
facilitate the experimental implementation of these algorithms in
the presence of noise. Moreover, a general MOTC framework has been presented
that extends beyond the specific problem to multiobservable control to
encompass more general quantum multiobjective optimization problems. Extensions to
problems involving objectives in multiple quantum systems \cite{Wolf2007} are straightforward.

The performance of MOTC algorithms has been compared to that of local gradient flow algorithms based on scalar
objective functions. Although the $\varepsilon(t)$-gradient flow
is always the locally optimal path, its projected path in $\U(N)$ is
generally much longer than those that can be tracked by global MOTC
algorithms. Even for single observable control problems, the latter
often require fewer iterations for convergence.

Features of the optimal control fields have been compared for various numbers of controlled observables.
Application of multiple observable expectation value constraints has been shown to generally increase the complexity
of optimal control fields on the level set of a quantum control landscape \cite{Raj2007} through the introduction of
higher frequency field modes beyond those present in the transition frequency spectrum of the internal Hamiltonian.

Two MOTC applications of particular importance are A) the simultaneous maximization of the expectation values
of a set of observables and B) the preparation of arbitrary mixed states. (A) corresponds to the problem
of identifying Pareto optimal solutions to the multiobservable control problem, i.e. control
fields $\e(t)$ such that all other fields have a lower value for at least one of the objective functions $\Phi_k$,
or else have the same value for all objectives. Whereas in generic multiobjective problems,
the Pareto frontier is very difficult to sample due to its irregular structure, the simple landscape topology
and geometry of quantum optimal control problems enables highly efficient methods for Pareto front exploration
based on MOTC.

Problem (B) requires the control of a large number of quantum observables (up to $N^2-1$,
for Hilbert space dimension $N$). For such tasks, the eigenvalue spectrum of the initial quantum state
plays an important role in determining the maximum possible optimization efficiency. In these cases, it
is usually possible to accelerate MOTC by  choosing an optimal set of observable operators to be measured
at each step. In this paper, we have focused on developing the formalism of MOTC and comparing the
algorithmic properties governing the efficiency of MOTC with those of gradient
flow and unitary matrix tracking algorithms. In a follow-up work, we will examine strategies for the efficient
experimental implementation of MOTC, as well as methods for combining quantum state estimation with MOTC for effective
Pareto front sampling.

\appendix

\section{Integration of kinematic gradient flows: $\rho(0)$ pure}

In this section, we analyze the trajectories followed by the quantum observable
maximization gradient flow (equation (11) in the main text) in order to shed light on the factors affecting its convergence rate.
This expression is cubic in $U$; however, through
the change of variables from $U_s(T)$ to $\rho_s(T) =
U_s(T)\rho(0)U^{\dag}_s(T)$, we can reexpress it as a quadratic
function:
\begin{eqnarray*}\label{rhoflow}
\frac{\p \rho_s(T)}{\p s} & =& -U_s(T)\rho(0)\frac{\p U^{\dag}_s(T)}{\p s} -\frac{\p U^{\dag}_s(T)}{\p s} \rho(0)U^{\dag}_s(T) \\
& = &\rho^2_s(T) \Theta - 2\rho_s(T) \Theta \rho_s(T) + \Theta \rho^2_s(T)\\
& =&\left[\rho_s(T),[\rho_s(T),\Theta ] \right]
\end{eqnarray*}
where $s$ denotes the algorithmic time variable of the gradient flow
in $\U(N)$. This quadratic expression for the gradient flow is in
so-called double bracket form
\cite{Brockett1991,Bloch1992,Helmke1994}.

Here, we provide an explicit formula pertaining to the analytical
solution to the above gradient flow for what is perhaps the most common objective in quantum optimal control theory and experiments, namely the maximization of the expectation value of an arbitrary observable
starting from a pure state $|i\rangle$.
Using the notation $|\psi_s(T) \rangle =U_s(T)|i\rangle$, the double bracket flow can in this case
be written:
$$
\frac{\p|\psi_s(T)\rangle}{\p s} =\frac{\p U_s(T)}{\p
s}|i\rangle=\left[ \Theta - \langle \psi_s(T)|\Theta|\psi_s(T)
\rangle I\right] ~ |\psi_s(T) \rangle.
$$
If we define $x(s) \equiv (|c_1(s)|^2,\ldots,|c_N(s)|^2)$, where
$c_1(s),\ldots,c_N(s)$ are the coordinates of $|\psi(s)\rangle$ in
the basis that diagonalizes $\Theta$, it can be verified that the
integrated gradient flow can be written:
\begin{eqnarray*}
x(s) &=& \frac{e^{2s\Theta}\cdot(|~c_1(0)|^2,\ldots,|~c_N(0)|^2)}{\sum_{k=1}^N|~c_k(0)|^2e^{2s\lambda_k}} \\
&=&\
\frac{(e^{2s\lambda_1}|~c_1(0)|^2,\ldots,e^{2s\lambda_N}|~c_N(0)|^2)}{\sum_{k=1}^N|~c_k(0)|^2e^{2s\lambda_k}}
\end{eqnarray*}
where $\lambda_1,\ldots, \lambda_N$ denote the eigenvalues of
$\Theta$.

The optimal solution to the search problem corresponds to the basis
vector $e_{j*}$ where $\Theta$ has its maximal eigenvalue (for
example, in the case of population transfer to a pure state
$|f\rangle$, $e_{j*} = |f\rangle$). The distance of the search
trajectory to the global optimum of the objective (framed on the
homogeneous space) can be expressed:
$$\|x(s)-e_{j*}\|^2 =
\|x(s)\|^2-2\frac{\langle e^{2s\Theta}x(0),e_{j*}\rangle }{\langle
e^{2s\Theta}x(0)\rangle }+1.$$
The time derivative of this distance function is
\begin{equation*}\label{phase}
\frac{\dd}{\dd t}\|x(s)-e_{j*}\|^2
=\frac{e^{2s\lambda_j*}x_{j*}(0)\sum_{k=1}^N(\lambda_{j*}-\lambda_k)e^{2s\lambda_k}x_k(0)}{\left[\sum_{k=1}^Ne^{2s\lambda_k}x_k(0)\right]^2}.
\end{equation*}

Solving for the zeroes of this time derivative reveals that the
distance between the current point on the search trajectory and the
solution can alternately increase and decrease with time.
Moreover, the same holds for the distance to the suboptimal critical points of the objective function.
As the rank and nondegeneracies in the eigenvalue spectrum of $\Theta$ increase, the
number of these suboptimal attractors of the search trajectory increases \cite{WuMike2008}, resulting in
an increase in the algorithmic path length.

\section{Natural basis functions for the dynamical gradient}

The expression for the dynamical observable maximization gradient,
Eq.(7) in the main text,
can be expanded as:
\begin{multline*}
\frac{\delta \Phi}{\delta \e(t)}
=-\frac{\imath}{\hbar}\tr\{\left[\Theta(T),\mu(t)\right]\rho(0)\}\\
=\frac {\imath}{\hbar} \sum_{i=1}^n p_i \sum_{j=1}^N \Big[\langle
i|\Theta(T)|j\rangle \langle j|\mu(t)|i\rangle +\\
-\langle i|\mu(t)|j\rangle \langle j|\Theta(T)|i\rangle \Big],
\end{multline*}

where the initial density matrix is given in terms of its nonzero eigenvalues
$p_i$ as $\rho(0)=\sum_{i=1}^n p_i |i\rangle\langle i|,~~ p_1 \geq
\ldots \geq p_n > 0, ~~\sum_{i=1}^n p_i = 1,$ where $n$ denotes the
rank of $\rho(0)$. The assumption of local surjectivity of the map
$\e(t) \mapsto U(T)$ implies that the functions $\langle
i|\mu(t)|j\rangle $ are $N^2$ linearly independent functions of
time. The functions
\begin{equation*}\label{gradbasis}
\langle i|\Theta(T)|j\rangle \langle j|\mu(t)|i\rangle -\langle
i|\mu(t)|j\rangle \langle j|\Theta(T)|i\rangle
\end{equation*}
therefore constitute natural basis functions for the gradient on the
domain $\e(t)$. From this the expression (18) in the main text for the dimension of the natural basis
follows.


\begin{thebibliography}{26}
\expandafter\ifx\csname natexlab\endcsname\relax\def\natexlab#1{#1}\fi
\expandafter\ifx\csname bibnamefont\endcsname\relax
  \def\bibnamefont#1{#1}\fi
\expandafter\ifx\csname bibfnamefont\endcsname\relax
  \def\bibfnamefont#1{#1}\fi
\expandafter\ifx\csname citenamefont\endcsname\relax
  \def\citenamefont#1{#1}\fi
\expandafter\ifx\csname url\endcsname\relax
  \def\url#1{\texttt{#1}}\fi
\expandafter\ifx\csname urlprefix\endcsname\relax\def\urlprefix{URL }\fi
\providecommand{\bibinfo}[2]{#2}
\providecommand{\eprint}[2][]{\url{#2}}

\bibitem[{\citenamefont{Assion et~al.}(1998)\citenamefont{Assion, Baumert,
  Bergt, Brixner, and Kiefer}}]{Assion1998}
\bibinfo{author}{\bibfnamefont{A.}~\bibnamefont{Assion}},
  \bibinfo{author}{\bibfnamefont{T.}~\bibnamefont{Baumert}},
  \bibinfo{author}{\bibfnamefont{M.}~\bibnamefont{Bergt}},
  \bibinfo{author}{\bibfnamefont{T.}~\bibnamefont{Brixner}}, \bibnamefont{and}
  \bibinfo{author}{\bibfnamefont{B.}~\bibnamefont{Kiefer}},
  \bibinfo{journal}{Science} \textbf{\bibinfo{volume}{282}},
  \bibinfo{pages}{5390} (\bibinfo{year}{1998}).

\bibitem[{\citenamefont{Ho and Rabitz}(2006)}]{HoRab2007a}
\bibinfo{author}{\bibfnamefont{T.-S.} \bibnamefont{Ho}} \bibnamefont{and}
  \bibinfo{author}{\bibfnamefont{H.}~\bibnamefont{Rabitz}},
  \bibinfo{journal}{J. Photochem. Photobiol. A} \textbf{\bibinfo{volume}{180}},
  \bibinfo{pages}{226} (\bibinfo{year}{2006}).

\bibitem[{\citenamefont{Rabitz et~al.}(2004)\citenamefont{Rabitz, Hsieh, and
  Rosenthal}}]{RabMik2004}
\bibinfo{author}{\bibfnamefont{H.}~\bibnamefont{Rabitz}},
  \bibinfo{author}{\bibfnamefont{M.}~\bibnamefont{Hsieh}}, \bibnamefont{and}
  \bibinfo{author}{\bibfnamefont{C.}~\bibnamefont{Rosenthal}},
  \bibinfo{journal}{Science} \textbf{\bibinfo{volume}{303}},
  \bibinfo{pages}{1998} (\bibinfo{year}{2004}).

\bibitem[{\citenamefont{Peirce et~al.}(1988)\citenamefont{Peirce, Dahleh, and
  Rabitz}}]{PeiDah1988}
\bibinfo{author}{\bibfnamefont{A.}~\bibnamefont{Peirce}},
  \bibinfo{author}{\bibfnamefont{M.}~\bibnamefont{Dahleh}}, \bibnamefont{and}
  \bibinfo{author}{\bibfnamefont{H.}~\bibnamefont{Rabitz}},
  \bibinfo{journal}{Phys. Rev. A} \textbf{\bibinfo{volume}{37}},
  \bibinfo{pages}{4950} (\bibinfo{year}{1988}).

\bibitem[{\citenamefont{Palao and Kosloff}(2002)}]{Kosloff2002}
\bibinfo{author}{\bibnamefont{Palao}} \bibnamefont{and}
  \bibinfo{author}{\bibfnamefont{R.}~\bibnamefont{Kosloff}},
  \bibinfo{journal}{Phys. Rev. Lett.} \textbf{\bibinfo{volume}{89}},
  \bibinfo{pages}{1883011} (\bibinfo{year}{2002}).

\bibitem[{\citenamefont{Palao and Kosloff}(2003)}]{Kosloff2003}
\bibinfo{author}{\bibfnamefont{J.}~\bibnamefont{Palao}} \bibnamefont{and}
  \bibinfo{author}{\bibfnamefont{R.}~\bibnamefont{Kosloff}},
  \bibinfo{journal}{Phys. Rev. A} \textbf{\bibinfo{volume}{68}},
  \bibinfo{pages}{062308} (\bibinfo{year}{2003}).

\bibitem[{\citenamefont{Weber et~al.}(2007)\citenamefont{Weber, Sauer,
  Plewicki, Merli, Wöste, and Lindinger}}]{Weber2007}
\bibinfo{author}{\bibfnamefont{S.~M.} \bibnamefont{Weber}},
  \bibinfo{author}{\bibfnamefont{F.}~\bibnamefont{Sauer}},
  \bibinfo{author}{\bibfnamefont{M.}~\bibnamefont{Plewicki}},
  \bibinfo{author}{\bibfnamefont{A.}~\bibnamefont{Merli}},
  \bibinfo{author}{\bibfnamefont{L.}~\bibnamefont{Wöste}}, \bibnamefont{and}
  \bibinfo{author}{\bibfnamefont{A.}~\bibnamefont{Lindinger}},
  \bibinfo{journal}{J. Mod. Opt.} \textbf{\bibinfo{volume}{54}},
  \bibinfo{pages}{2659} (\bibinfo{year}{2007}).

\bibitem[{\citenamefont{Bonacina et~al.}(2007)\citenamefont{Bonacina,
  Extermann, Rondi, Boutou, and Wolf}}]{Wolf2007}
\bibinfo{author}{\bibfnamefont{L.}~\bibnamefont{Bonacina}},
  \bibinfo{author}{\bibfnamefont{J.}~\bibnamefont{Extermann}},
  \bibinfo{author}{\bibfnamefont{A.}~\bibnamefont{Rondi}},
  \bibinfo{author}{\bibfnamefont{V.}~\bibnamefont{Boutou}}, \bibnamefont{and}
  \bibinfo{author}{\bibfnamefont{J.}~\bibnamefont{Wolf}},
  \bibinfo{journal}{Phys. Rev. A} \textbf{\bibinfo{volume}{76}},
  \bibinfo{pages}{023408} (\bibinfo{year}{2007}).

\bibitem[{\citenamefont{Levis et~al.}(2001)\citenamefont{Levis, Menkir, and
  Rabitz}}]{Levis2001}
\bibinfo{author}{\bibfnamefont{R.}~\bibnamefont{Levis}},
  \bibinfo{author}{\bibfnamefont{G.}~\bibnamefont{Menkir}}, \bibnamefont{and}
  \bibinfo{author}{\bibfnamefont{H.}~\bibnamefont{Rabitz}},
  \bibinfo{journal}{Science} \textbf{\bibinfo{volume}{292}},
  \bibinfo{pages}{709} (\bibinfo{year}{2001}).

\bibitem[{\citenamefont{Deb et~al.}(2002)\citenamefont{Deb, Pratap, Agarwal,
  and Meyarivan}}]{Deb2002}
\bibinfo{author}{\bibfnamefont{K.}~\bibnamefont{Deb}},
  \bibinfo{author}{\bibfnamefont{A.}~\bibnamefont{Pratap}},
  \bibinfo{author}{\bibfnamefont{S.}~\bibnamefont{Agarwal}}, \bibnamefont{and}
  \bibinfo{author}{\bibfnamefont{T.}~\bibnamefont{Meyarivan}},
  \bibinfo{journal}{IEEE Trans. Evol. Comp.} \textbf{\bibinfo{volume}{6}},
  \bibinfo{pages}{182} (\bibinfo{year}{2002}).

\bibitem[{\citenamefont{Chakrabarti and Rabitz}(2007)}]{Raj2007}
\bibinfo{author}{\bibfnamefont{R.}~\bibnamefont{Chakrabarti}} \bibnamefont{and}
  \bibinfo{author}{\bibfnamefont{H.}~\bibnamefont{Rabitz}},
  \bibinfo{journal}{Int. Rev. Phys. Chem.} \textbf{\bibinfo{volume}{26}},
  \bibinfo{pages}{671} (\bibinfo{year}{2007}).

\bibitem[{\citenamefont{Wu et~al.}(2008{\natexlab{a}})\citenamefont{Wu, Pechen,
  Rabitz, Hsieh, and Tsou}}]{WuPech2008}
\bibinfo{author}{\bibfnamefont{R.}~\bibnamefont{Wu}},
  \bibinfo{author}{\bibfnamefont{A.}~\bibnamefont{Pechen}},
  \bibinfo{author}{\bibfnamefont{H.}~\bibnamefont{Rabitz}},
  \bibinfo{author}{\bibfnamefont{M.}~\bibnamefont{Hsieh}}, \bibnamefont{and}
  \bibinfo{author}{\bibfnamefont{B.}~\bibnamefont{Tsou}}, \bibinfo{journal}{J.
  Math. Phys.} \textbf{\bibinfo{volume}{49}}, \bibinfo{pages}{022108}
  (\bibinfo{year}{2008}{\natexlab{a}}).

\bibitem[{\citenamefont{Roslund et~al.}(2008)\citenamefont{Roslund, Roth, and
  Rabitz}}]{Roslund2007}
\bibinfo{author}{\bibfnamefont{J.}~\bibnamefont{Roslund}},
  \bibinfo{author}{\bibfnamefont{M.}~\bibnamefont{Roth}}, \bibnamefont{and}
  \bibinfo{author}{\bibfnamefont{H.}~\bibnamefont{Rabitz}},
  \bibinfo{journal}{in preparation}  (\bibinfo{year}{2008}).

\bibitem[{\citenamefont{Hillermeier}(2001)}]{Hillermeier2001}
\bibinfo{author}{\bibfnamefont{C.}~\bibnamefont{Hillermeier}},
  \emph{\bibinfo{title}{Nonlinear multiobjective optimization: a generalized
  homotopy approach}} (\bibinfo{publisher}{Birkhauser},
  \bibinfo{address}{Basel}, \bibinfo{year}{2001}).

\bibitem[{\citenamefont{Rabitz et~al.}(2006)\citenamefont{Rabitz, Hsieh, and
  Rosenthal}}]{Mike2006a}
\bibinfo{author}{\bibfnamefont{H.}~\bibnamefont{Rabitz}},
  \bibinfo{author}{\bibfnamefont{M.}~\bibnamefont{Hsieh}}, \bibnamefont{and}
  \bibinfo{author}{\bibfnamefont{C.}~\bibnamefont{Rosenthal}},
  \bibinfo{journal}{J. Chem. Phys.} \textbf{\bibinfo{volume}{124}},
  \bibinfo{pages}{204107} (\bibinfo{year}{2006}).

\bibitem[{\citenamefont{Wu et~al.}(2008{\natexlab{b}})\citenamefont{Wu, Rabitz,
  and Hsieh}}]{WuMike2008}
\bibinfo{author}{\bibfnamefont{R.}~\bibnamefont{Wu}},
  \bibinfo{author}{\bibfnamefont{H.}~\bibnamefont{Rabitz}}, \bibnamefont{and}
  \bibinfo{author}{\bibfnamefont{M.}~\bibnamefont{Hsieh}}, \bibinfo{journal}{J.
  Phys. A} \textbf{\bibinfo{volume}{41}}, \bibinfo{pages}{015006}
  (\bibinfo{year}{2008}{\natexlab{b}}).

\bibitem[{\citenamefont{Bertero et~al.}(1985)\citenamefont{Bertero, Mol, and
  Pike}}]{Bertero1985}
\bibinfo{author}{\bibfnamefont{M.}~\bibnamefont{Bertero}},
  \bibinfo{author}{\bibfnamefont{C.~D.} \bibnamefont{Mol}}, \bibnamefont{and}
  \bibinfo{author}{\bibfnamefont{E.~R.} \bibnamefont{Pike}},
  \bibinfo{journal}{Inverse Probl.} \textbf{\bibinfo{volume}{1}},
  \bibinfo{pages}{301} (\bibinfo{year}{1985}).

\bibitem[{\citenamefont{Dominy and Rabitz}(2008)}]{Dominy2008}
\bibinfo{author}{\bibfnamefont{J.}~\bibnamefont{Dominy}} \bibnamefont{and}
  \bibinfo{author}{\bibfnamefont{H.}~\bibnamefont{Rabitz}},
  \bibinfo{journal}{J. Phys. A, in press}  (\bibinfo{year}{2008}).

\bibitem[{\citenamefont{Bonnard and Chyba}(2003)}]{Bonnard2003}
\bibinfo{author}{\bibfnamefont{B.}~\bibnamefont{Bonnard}} \bibnamefont{and}
  \bibinfo{author}{\bibfnamefont{M.}~\bibnamefont{Chyba}},
  \emph{\bibinfo{title}{Singular trajectories and their role in control
  theory}} (\bibinfo{publisher}{Springer}, \bibinfo{address}{Berlin},
  \bibinfo{year}{2003}).

\bibitem[{\citenamefont{Brockett}(1991)}]{Brockett1991}
\bibinfo{author}{\bibfnamefont{R.}~\bibnamefont{Brockett}},
  \bibinfo{journal}{Linear Alg Appl.} \textbf{\bibinfo{volume}{146}},
  \bibinfo{pages}{79} (\bibinfo{year}{1991}).

\bibitem[{\citenamefont{Rothman et~al.}(2005)\citenamefont{Rothman, Ho, and
  Rabitz}}]{Rothman2005}
\bibinfo{author}{\bibfnamefont{A.}~\bibnamefont{Rothman}},
  \bibinfo{author}{\bibfnamefont{T.}~\bibnamefont{Ho}}, \bibnamefont{and}
  \bibinfo{author}{\bibfnamefont{H.}~\bibnamefont{Rabitz}},
  \bibinfo{journal}{Phys. Rev. A} \textbf{\bibinfo{volume}{72}},
  \bibinfo{pages}{023416} (\bibinfo{year}{2005}).

\bibitem[{\citenamefont{Uhlmann}(1976)}]{Uhlmann1976}
\bibinfo{author}{\bibfnamefont{A.}~\bibnamefont{Uhlmann}},
  \bibinfo{journal}{Rep. Math. Phys.} \textbf{\bibinfo{volume}{9}},
  \bibinfo{pages}{273} (\bibinfo{year}{1976}).

\bibitem[{\citenamefont{Yip et~al.}(2003)\citenamefont{Yip, Mazziotti, and
  Rabitz}}]{Yip2003}
\bibinfo{author}{\bibfnamefont{F.}~\bibnamefont{Yip}},
  \bibinfo{author}{\bibfnamefont{D.}~\bibnamefont{Mazziotti}},
  \bibnamefont{and} \bibinfo{author}{\bibfnamefont{H.}~\bibnamefont{Rabitz}},
  \bibinfo{journal}{J. Chem. Phys.} \textbf{\bibinfo{volume}{118}},
  \bibinfo{pages}{8168} (\bibinfo{year}{2003}).

\bibitem[{\citenamefont{Press et~al.}(2002)\citenamefont{Press, Teukolsky,
  Vetterling, and Flannery}}]{Press2002}
\bibinfo{author}{\bibfnamefont{W.}~\bibnamefont{Press}},
  \bibinfo{author}{\bibfnamefont{S.}~\bibnamefont{Teukolsky}},
  \bibinfo{author}{\bibfnamefont{W.}~\bibnamefont{Vetterling}},
  \bibnamefont{and} \bibinfo{author}{\bibfnamefont{B.}~\bibnamefont{Flannery}},
  \emph{\bibinfo{title}{Numerical recipes in $C^{++}$}}
  (\bibinfo{publisher}{Cambridge Univ. Press}, \bibinfo{address}{Cambridge},
  \bibinfo{year}{2002}).

\bibitem[{\citenamefont{Bloch et~al.}(1992)\citenamefont{Bloch, Brockett, and
  Ratiu}}]{Bloch1992}
\bibinfo{author}{\bibfnamefont{A.}~\bibnamefont{Bloch}},
  \bibinfo{author}{\bibfnamefont{R.}~\bibnamefont{Brockett}}, \bibnamefont{and}
  \bibinfo{author}{\bibfnamefont{T.}~\bibnamefont{Ratiu}},
  \bibinfo{journal}{Commun. Math. Phys.} \textbf{\bibinfo{volume}{147}},
  \bibinfo{pages}{57} (\bibinfo{year}{1992}).

\bibitem[{\citenamefont{Helmke and Moore}(1994)}]{Helmke1994}
\bibinfo{author}{\bibfnamefont{U.}~\bibnamefont{Helmke}} \bibnamefont{and}
  \bibinfo{author}{\bibfnamefont{J.}~\bibnamefont{Moore}},
  \emph{\bibinfo{title}{Optimization and dynamical systems}}
  (\bibinfo{publisher}{Springer-Verlag}, \bibinfo{address}{London},
  \bibinfo{year}{1994}).

\end{thebibliography}

\end{document}